\documentclass[nobibnotes,nofootinbib,aps,pre,showpacs, preprint]{revtex4} % for 2 columns style with line numbers
\usepackage{color}%,graphics}
\usepackage{setspace}
\usepackage{graphics}
\usepackage{graphicx}
\usepackage{color}
\usepackage{amsmath}
\usepackage{amssymb}
\usepackage{bm}
\usepackage{float}
\usepackage{epsfig}
\usepackage{epstopdf}
\usepackage{color,graphics}
\usepackage[toc,page]{appendix}
\usepackage{array,multirow,makecell}
\usepackage{amsfonts}
\begin{document}

\title{Bernstein-Greene-Kruskal approach for the quantum Vlasov equation}
%\shorttitle{xxx} %Insert here a short version of the title if it exceeds 70 characters

\author{F. Haas}
%\shortauthor{F. Author \etal}

\affiliation{                    
Instituto de F\'{\i}sica, Universidade Federal do Rio Grande do Sul, Av. Bento Gon\c{c}alves 9500, 91501-970 Porto Alegre, RS, Brasil %\\
 % \inst{2} Second Institute - Address
}

\begin{abstract}
The one-dimensional stationary quantum Vlasov equation is analyzed using the energy as one of the dynamical variables, 
similarly as in the solution of the Vlasov-Poisson system by means of the Bernstein-Greene-Kruskal method. In the semiclassical case where quantum tunneling effects are small, an infinite series solution is developed and shown to be immediately integrable up to a recursive chain of quadratures in position space only. { As it stands, the treatment of the self-consistent, Wigner-Poisson system is beyond the scope of the method, which assumes} a given smooth { time-independent} external potential. Accuracy tests for the series expansion are also provided. Examples of anharmonic potentials are worked out up to a high order on the quantum diffraction parameter.
\end{abstract}


\maketitle

\section{Introduction}

The Wigner function was introduced almost one century ago \cite{Wigner}, as a distinguished joint probability distribution in quantum mechanics. The applications of the Wigner function appear in many contexts, such as quantum entanglement, classical and quantum information processing, quantum electronics and quantum chemistry, see \cite{Weinbub} for a recent review. 
Specifically in quantum plasmas, it plays a r\^ole for instance in nonlinear waves and wavebreaking \cite{Kull}, Landau damping effects on  bremsstrahlung process \cite{Jung}, quantum free-electron-lasers \cite{Boni} and the bound states near a moving charge, using Lindhard's dielectric function which can be derived from the Wigner-Poisson system \cite{Else}. It is therefore an important issue, to derive accurate expressions of the Wigner function, which is the subject of the present work.  

The Wigner function obeys the so-called Wigner-Moyal or quantum Vlasov equation \cite{Jungel2009}. The classical limit of the Wigner-Moyal equation is the Vlasov equation, which is solved by an arbitrary function of the constants of
motion of the system. In the time-invariant case, this allows the derivation of Bernstein-Greene-Kruskal equilibria \cite{Bernstein} for the Vlasov-Poisson system of the classical plasma, taking the energy as the central dynamical variable. However, as expected, the quantum kinetic equation does not { preserve} the classical constants of motion. In view of this, most approaches for the Wigner-Moyal equation rest on semiclassical treatments, restricted to the first order quantum correction \cite{Luque}-\cite{Lange}. This includes the original article by Wigner \cite{Wigner}, where the lowest order quantum correction to the Maxwell-Boltzmann equilibrium was evaluated. Nevertheless, already in \cite{Wigner} the possibility of series solutions up to arbitrary order has been proposed, see also \cite{Bose}. In addition, the role of the energy as an useful dynamical variable has been identified, for a certain class of solutions of the stationary one-dimensional Wigner-Moyal equation and Wigner-Poisson system not restricted to the semiclassical case \cite{HS}. The connection between the quantum mechanical and classical time-evolutions by means of a dynamical (Ermakov) invariant was also recognized \cite{Schuch}. However, the series expansion solution jointly with the choice of the energy as central object was not proposed before. In this way we will show that not only an infinite chain of partial differential equations is found \cite{Wigner}. Much differently, we are allowed to immediate quadrature in position space, recursively and up to arbitrary order on the quantum parameter.  { By definition, the treatment of the self-consistent, Wigner-Poisson system is beyond the scope of the method, at least in its present form, which assumes a given smooth time-independent external potential.}

The purpose of the present work is to demonstrate the usefulness of the energy as a key dynamical variable in the solution of the one-dimensional quantum Vlasov equation, in terms of a power series solution which can be easily implemented up to arbitrary order on the quantum effects, reducing the problem to quadrature in configuration space only. The approach can be viewed as the quantum analogue of the Bernstein-Greene-Kruskal method \cite{Bernstein}. However, our treatment does not consider a self-consistent field, as would be necessary for quantum plasmas for instance, because in this situation both the Wigner function and the scalar field would be necessarily expanded. Likewise, the case involving electromagnetic fields treated in a gauge invariant manner will be left for future works \cite{NJP, Nedjalkov}. It must be emphasized that the details of the zeroth-order solution (the classical limit) are not decisive for the procedure. 

This work is organized as follows. We introduce the one-dimensional quantum Vlasov equation written in dimensionless variables appropriate for semiclassical expansions. Afterward we consider the time-independent situation and a transformation of variables where a key 
{ role} is played by the classical Hamiltonian. The power series solution on the scaled quantum parameter is shown to be always reducible to a chain of quadratures, once the external potential is specified, leaving free the classical limit of the Wigner function. The recursive procedure is illustrated in the cases of quartic potentials and of a modulated harmonic { potential}, yielding the Wigner function up to high order on the { quantum-tunneling-effects} parameter. Finally, our conclusions are outlined.

\section{Statement of the problem}

The quantum Vlasov equation, or Wigner-Moyal equation, is the kinetic equation for the evolution of the Wigner quasi-probability distribution function \cite{Jungel2009}. In one spatial dimension, it reads
\begin{equation}
\label{e1}
\frac{\partial f}{\partial t} + \frac{p}{m}\,\frac{\partial f}{\partial q} - \theta_{\hbar}[V]f = 0 \,, \quad q, p \in \Re \,, t > 0 \,,
\end{equation}
where $f = f(q,p,t)$ is the Wigner function, $m$ is the mass, $\hbar$ is the reduced Planck constant and $V = V(q,t)$ is the potential. The quantity $\theta_{\hbar}[V]$ is a pseudo-differential operator \cite{Taylor} defined in terms of the symbol 
\begin{equation}
\label{e2}
(\delta V)_{\hbar}(q,\eta,t) \equiv \frac{i}{\hbar}\left(V(q + \frac{\hbar\eta}{2},t) - V(q - \frac{\hbar\eta}{2},t)\right) \,,
\end{equation}
{\it videlicet}, 
\begin{eqnarray}
\label{e3}
&\strut& (\theta_{\hbar}[V]f)(q,p,t) =  \\ 
&=& \frac{1}{2\pi}\int_{\Re}\int_{\Re} (\delta V)_{\hbar}(q,\eta,t) f(q,p',t) e^{i(p-p')\eta}dp' d\eta \,, \nonumber
\end{eqnarray}
assuming appropriate functions $f, V$.

In the semiclassical limit $\hbar \rightarrow 0$ detailed e.g. in \cite{Markowich}, Eq. (\ref{e1}) becomes the one-dimensional Vlasov equation,
\begin{equation}
\label{e4}
\frac{\partial f}{\partial t} + \frac{p}{m}\,\frac{\partial f}{\partial q} - \frac{\partial V}{\partial q}\frac{\partial f}{\partial p} = 0 \,.
\end{equation}

Expanding to higher orders yields 
\begin{eqnarray}
\frac{\partial f}{\partial t} \! &+& \! \frac{p}{m}\,\frac{\partial f}{\partial q} - \frac{\partial V}{\partial q}\frac{\partial f}{\partial p} 
+ \nonumber \\ \label{e5}
 &+& \frac{\hbar^2}{24}\frac{\partial^3 V}{\partial q^3}\frac{\partial^3 f}{\partial p^3}
- \frac{\hbar^4}{1920}\frac{\partial^5 V}{\partial q^5}\frac{\partial^5 f}{\partial p^5}
= {\cal O}(\hbar^6) \,.
\end{eqnarray}
It is evident that the form of the quantum correction terms makes it difficult to find an exact solution of the quantum Vlasov equation, even in the stationary case. The exception if for a quadratic potential, where the series expansion quickly terminates so that the quantum and classical Vlasov equations coincide.

It is convenient to adopt scaled dimensionless variables defined by  
\begin{equation}
\label{e6}
q_s = \frac{q}{q_0} \,,  p_s = \frac{p}{p_0} \,, f_s = \frac{f}{p_0 q_0} \,,  V_s = \frac{V}{V_0} \,,  \hbar_s = \frac{\hbar}{p_0 q_0} \,,
\end{equation}
where $q_0$, $p_0$ and $V_0 = p_0^2/m$ are respectively characteristics length, momentum and energy. 
{ For applications, the natural scaling sets the parameters of the external potential to unity, as much as possible. Just for the sake of illustration, for the harmonic potential $V = m \omega^2 q^2/2$ with angular frequency $\omega$ it is reasonable to set $V_s = q_s^2/2$, implying $p_0 = m \omega q_0$. In this case, one has $\hbar_s = \hbar\omega/(p_0^2/m)$, where $V_0 = p_0^2/m$ can be estimated by the thermal or Fermi energies, according to the degeneracy degree.  This prescription should be adapted to each physical system, and determines the concrete value of $\hbar_s$ therein. In passing, we note that $\hbar_s$ is a measure of the quantum-diffraction effects (or wave effects in general), in the sense that in the classical limit $\hbar_s \rightarrow 0$ the Wigner-Moyal equation reduces to the Vlasov equation.}

Dropping the subscript $s$, one has the stationary ($\partial/\partial t = 0$) quantum Vlasov equation 
\begin{equation}
\label{e7}
p\,\frac{\partial f}{\partial q} - \theta_{\hbar}[V]f = 0 \,, \quad q, p \in \Re \,,
\end{equation}
where $f = f(q,p)$, $V = V(q)$, %the pseudo-diferential operator xxx
\begin{equation}
\label{e8}
(\delta V)_{\hbar}(q,\eta) = \frac{i}{\hbar}\left(V(q + \frac{\hbar\eta}{2}) - V(q - \frac{\hbar\eta}{2})\right) \,,
\end{equation}
and
\begin{eqnarray}
\label{e9}
&\strut& (\theta_{\hbar}[V]f)(q,p) = \\ &=& \frac{1}{2\pi}\int_{\Re}\int_{\Re} (\delta V)_{\hbar}(q,\eta) f(q,p') e^{i(p-p')\eta}dp' d\eta \,. \nonumber
\end{eqnarray}
The rescaling provides a more sensible approach for the semiclassical limit, in terms of a series expansion on the dimensionless $\hbar$ parameter. In the following, we derive a concise expression for the formal series solution of Eq. { (\ref{e7})} up to arbitrary order on the { quantum-tunneling} effects, for arbitrary zeroth-order Wigner function in the classical limit.

\section{Formal power series solution. Recursion formula. Validity conditions}

In the classical limit, it is known that the Vlasov equation is solved for an arbitrary function of the constants of motion (Jeans theorem). In order to take advantage of this, it is appropriate to define the new variables $(x,H)$ according to 
\begin{equation}
\label{e10}
x = q \,, \quad H = \frac{p^2}{2} + V(q) \,, 
\end{equation}
so that 
\begin{equation}
\label{e11}
\frac{\partial}{\partial q} = \frac{\partial V}{\partial x}\frac{\partial}{\partial H} + \frac{\partial}{\partial x} \,, \quad \frac{\partial}{\partial p} = p\frac{\partial}{\partial H} \,.
\end{equation}

Equation (7) for $f = f(x,H)$ becomes
\begin{equation}
\label{ee}
p\frac{\partial f}{\partial x} = \sum_{j=1}^{\infty} \frac{1}{(2j+1)!}\left(\frac{i\hbar}{2}\right)^{2j}\frac{\partial^{2j+1}V(x)}{\partial x^{2j+1}}\left(\frac{\partial}{\partial p}\right)^{2j+1}f \,,
\end{equation}
where at this stage the momentum $p$ is maintained. Notice that the right-hand side of Eq. (\ref{ee}) is formally zero in the classical limit. The potential and the Wigner function are assumed to be smooth, otherwise the present treatment does not apply.

A direct calculation shows that
\begin{equation}
\label{e12}
\frac{\partial^j\!f}{\partial p^j} = \sum_{k=0}^{\infty} \frac{j!\,p^{j-2k}}{2^k\,k!\,(j-2k)!} \frac{\partial^{j-k}f}{\partial H^{j-k}} \,.
\end{equation}
In passing, for the interpretation of Eq. (\ref{e12}) we note that $0! = 1$ and $1/j! = 0$ if $j$ is a negative integer. Equation (\ref{e12}) allows to convert Eq. (\ref{ee}) into 
\begin{eqnarray}
\label{e13}
\frac{\partial f}{\partial x} &=& \sum_{j=1}^{\infty}\left(\frac{i\hbar}{2}\right)^{2j}\frac{\partial^{2j+1}V(x)}{\partial x^{2j+1}} \times \\ &\times& \sum_{k=0}^{j}\frac{1}{(2j-2k+1)!}\frac{p^{2(j-k)}}{2^k\,k!}\frac{\partial^{2j-k+1}\,f}{\partial H^{2j-k+1}} \,. \nonumber 
\end{eqnarray}
Using $p^2 = 2 (H-V(x))$ finally yields
\begin{eqnarray}
\label{e14}
\frac{\partial f}{\partial x} &=& \sum_{j=1}^{\infty}\left(-\,\frac{\hbar^2}{2}\right)^{j}\frac{\partial^{2j+1}V(x)}{\partial x^{2j+1}} \times \\ &\times& \sum_{k=0}^{j}\frac{(H-V(x))^{j-k}}{2^{2k}\,k!\,(2j-2k+1)!}\frac{\partial^{2j-k+1}\,f}{\partial H^{2j-k+1}} \,. \nonumber 
\end{eqnarray}

It is natural to seek for a series solution
\begin{equation}
\label{e15}
f = \sum_{j=0}^{\infty}\hbar^{2j}f_j(x,H) \,,
\end{equation}
provided the scaled Planck constant is a small parameter. Inserting into Eq. (\ref{e14}), to zero order one has 
\begin{equation}
\label{e16}
\frac{\partial f_0}{\partial x} = 0 \quad \Rightarrow \quad f_0 = f_0(H) \,.
\end{equation}
In the classical limit the stationary Wigner function depends on the energy only, as expected. 

The next order correction can be expressed as
\begin{eqnarray}
\frac{\partial f_1}{\partial x} = - \,\frac{\partial}{\partial x}\Bigl[\!\!\!\!\!\!\!&\strut&\!\!\!\!\!\!\! \frac{1}{2}\frac{\partial^2 V}{\partial x^2}\left(\frac{(H - V)}{6}\frac{\partial^3 f_0}{\partial H^3} + \frac{1}{4}\frac{\partial^2 f_0}{\partial H^2}\right) +\nonumber \\ \label{e17}
 &+& \frac{1}{24}\left(\frac{\partial V}{\partial x}\right)^2 \frac{\partial^3 f_0}{\partial H^3}\Bigr] \,,
\end{eqnarray}
yielding 
\begin{eqnarray}
\nonumber 
f_1 = &-&\,\frac{1}{2}\frac{\partial^2 V}{\partial x^2}\left(\frac{(H - V)}{6}\frac{\partial^3 f_0}{\partial H^3} + \frac{1}{4}\frac{\partial^2 f_0}{\partial H^2}\right) \\ &-&\, \frac{1}{24}\left(\frac{\partial V}{\partial x}\right)^2 \frac{\partial^3 f_0}{\partial H^3} + f_{01}(H) \,, 
\label{e18}
\end{eqnarray}
where $f_{01}(H)$ is an arbitrary function of  $H$. It can be verified that if the potential is quadratic, then $f_1$ becomes a function of $H$ only, 
\begin{eqnarray}
V &=& a + b x + c x^2 \\ &\Rightarrow& f_1 = \frac{1}{24}(4ac - b^2 { - 4cH})\frac{\partial^3 f_0}{\partial H^3} - \frac{c}{4}\frac{\partial^2 f_0}{\partial H^2} + f_{01}(H) \,, \nonumber 
\end{eqnarray}
which is expected since in this case the quantum corrections to the Vlasov equation disappear (here $a, b, c$ are constants). The same holds for the higher order corrections when the potential is quadratic. 

When $f_0$ is a Maxwellian, Eq. (\ref{e18}) reproduces the Wigner result for a quantum corrected thermodynamic equilibrium \cite{Wigner}. However, the expression (\ref{e18}) holds for arbitrary $f_0$, for instance for Fermi-Dirac or Bose-Einstein equilibria and beyond. 

From Eqs. (\ref{e14}) and (\ref{e15}), to general order one derives
\begin{eqnarray}
\label{e19}
\frac{\partial f_l}{\partial x} \!\!\!\!\!\! &\strut& \!\!\!\!\!\! = \sum_{j=1}^{l} \left(-\,\frac{1}{2}\right)^{j}\frac{\partial^{2j+1}V}{\partial x^{2j+1}} \times \\ &\times&\sum_{k=0}^{j}\frac{(H-V)^{j-k}}{2^{2k}k!(2j-2k+1)!}\frac{\partial^{2j-k+1}f_{l-j}}{\partial H^{2j-k+1}} \,,\,\, l = 0,1,\dots \nonumber
\end{eqnarray}
Although it can be cumbersome to find the general expression for $f_l, \,l \geq 2$, for a specific $V(x)$ the higher order corrections are directly obtained by quadrature of the right-hand side of Eq. (\ref{e19}), wherein $H$ is just a parameter. Indeed, $f_2$ will be found inserting $f_{0,1}$ from Eqs. (\ref{e16}) and (\ref{e18}) together with the external potential and after a quadrature. Similarly for $f_3$ which {\it will depend} on $f_{0,1,2}$, and so on in an infinite recursive chain of quadratures in position space only.  For this reason, the details of $f_0(H)$ (the classical equilibrium) are obviously not decisive for the step-by-step procedure. In other words, instead of a cumbersome sequence of partial differential equations to be solved order by order for the quantum corrections, one finds a sequence or first-order ordinary differential equations all reducible to quadratures. The calculation is easily implemented with a { computer-algebra} program. 

It is evident from Eq. (\ref{e19}) that all $f_j$ are defined up to the addition of an arbitrary function of $H$. For instance, if one starts with $f_0 \equiv 0$, one gets $f_1 = f_{01}(H)$ and then from Eq. (\ref{e19}) the next order result is
\begin{eqnarray}
\nonumber f_2 \!\!\!\!\!\! &\strut& \!\!\!\!\!\! =  -\,\frac{1}{2}\frac{\partial^2 V}{\partial x^2}\left(\frac{(H - V)}{6}\frac{\partial^3 
f_{01}}{\partial H^3} + \frac{1}{4}\frac{\partial^2 f_{01}}{\partial H^2}\right) \\ &-& \frac{1}{24}\left(\frac{\partial V}{\partial x}\right)^2 \frac{\partial^3 f_{01}}{\partial H^3} + f_{02}(H) \,,
\label{e20}
\end{eqnarray} 
where $f_{02}(H)$ is an arbitrary function. 
In this case one has $f/\hbar^2 = f_{01}(H) + \hbar^2 f_2 + \dots\,,$ exactly reproducing Eq. (\ref{e18}) where $f_0(H) \neq 0$, with the replacements $f_0(H) \rightarrow f_{01}(H),\, f_{01}(H) \rightarrow f_{02}(H)$, as seen by comparison. 
In general, it can be directly shown that
\begin{eqnarray}
\label{e21}
f = \!\!\!\!\!\!&\strut&\!\!\!\!\!\!\!\!\! f_0(H) + \hbar^2 F_{1}(f_0(H)) + \hbar^4 F_2(f_0(H)) + \dots  \\
&+& \hbar^2 \left[f_{01}(H) + \hbar^2 F_{1}(f_{01}(H)) + \hbar^4 F_2(f_{01}(H)) + \dots \right] \nonumber \\
&+& \hbar^4 \left[f_{02}(H) + \hbar^2 F_{1}(f_{02}(H)) + \hbar^4 F_2(f_{02}(H)) + \dots \right] \,, \nonumber 
\end{eqnarray}
where the $F_j$ are linear operators such that $F_j(0) = 0\,,\, j = 1, 2, \dots$  In this context each bracket term in Eq. (\ref{e21}) starting with a different seed function $f_{0j}(H)$ separately corresponds to a solution of the quantum Vlasov equation, which is linear in the case of an external potential. With this proviso we can omit the arbitrary functions, setting $f_{0j} = 0$ and focusing on the determination of $F_{1,2,...}$, which simplifies the algebra together with a saving of computer running time. It is interesting to note that the structure of the solutions of the stationary one-dimensional quantum Vlasov equation contains a certain arbitrary functional dependence on the energy first integral, as much as in the classical case. 

It is difficult to determine the convergence of the series expansion. However, there are some necessary conditions for a faithful Wigner function, which should correspond to a positive definite density matrix \cite{Hillery}, namely, 
\begin{eqnarray}
P_q(q,\hbar) &=& \frac{\int_{\Re} dp\,f}{\int_{\Re} dp \int_{\Re} dq \,f}  \geq 0 \,,
\label{x1} \\
P_p(p,\hbar) &=& \frac{\int_{\Re} dq\,f}{\int_{\Re} dp \int_{\Re} dq \,f} \geq 0 \,,
\label{x2} \\
Q(\hbar) &=& \frac{\int_{\Re} dp \int_{\Re} dq \,f^2}{(\int_{\Re} dp \int_{\Re} dq \,f)^2} \leq \frac{1}{2\pi\hbar}  \,,
\label{x3}
\end{eqnarray}
valid for arbitrary normalization. 
Equations (\ref{x1}) and (\ref{x2}) correspond to positive definite marginal probability distributions in position and momentum spaces. Equation (\ref{x3}) rules out too spiky Wigner functions violating the uncertainty principle. These necessary conditions provide an useful test for the accuracy of the series solution.

\section{Example: one-dimensional Goldstone potential}

For the sake of illustration, consider the symmetry breaking one-dimensional Goldstone potential 
\begin{equation}
\label{e22}
V = - \frac{q^2}{2} + \frac{q^4}{4} \,,
\end{equation}
in rescaled variables. Being the simplest symmetric model besides the quadratic potential so that the Vlasov and quantum Vlasov equations do not coincide, the quartic oscillator was investigated in the context of quantum echoes \cite{Manfredi}.   We carried on the series in Eq. (\ref{e15}) up to ${\cal O}(\hbar^{10})$, solving the chain of equations shown in Eq. (\ref{e15}) to the same order, always setting the additive functions of $H$ to zero, having in mind the structure detected in Eq. (\ref{e21}). 

For instance, using the computer algebra software Wolfram Mathematica 11.0 it is easy to quickly derive
\begin{eqnarray}
f_{1}(x,H) &=& \frac{1}{48} \Bigl(\left(6-18 x^2\right) f_0^{(2)}+ \nonumber \\ &+& \left(4 H-12 H x^2-3 x^4+x^6\right) f_0^{(3)}\Bigr) \,, \\
f_{2}(x,H) &=& \frac{x^2}{4608} \Bigl[252 \left(-2+3 x^2\right) f_0^{(4)} \nonumber \\ &-& 18 \left(32 H+(6-48 H) x^2-16 x^4+5 x^6\right) f_0^{(5)}\nonumber \\ &+&\Bigl(-96 H^2+24
H (-1+6 H) x^2+\nonumber \\ &+& 80 H x^4+(9-24 H) x^6-6 x^8+x^{10}\Bigr) f_0^{(6)}\Bigr] \,,\nonumber \\ &\strut& 
\end{eqnarray}
which yields the ${\cal O}(\hbar^{4})$ correction, obviously valid for arbitrary seed function $f_0$, denoting derivatives as $f_{0}^{(j)} = \partial^j\!f_{0}/\partial H^j$. The heavy expressions for the next order corrections will be omitted.  

\begin{figure}[h]
\begin{center}
\includegraphics[width=4.5in]{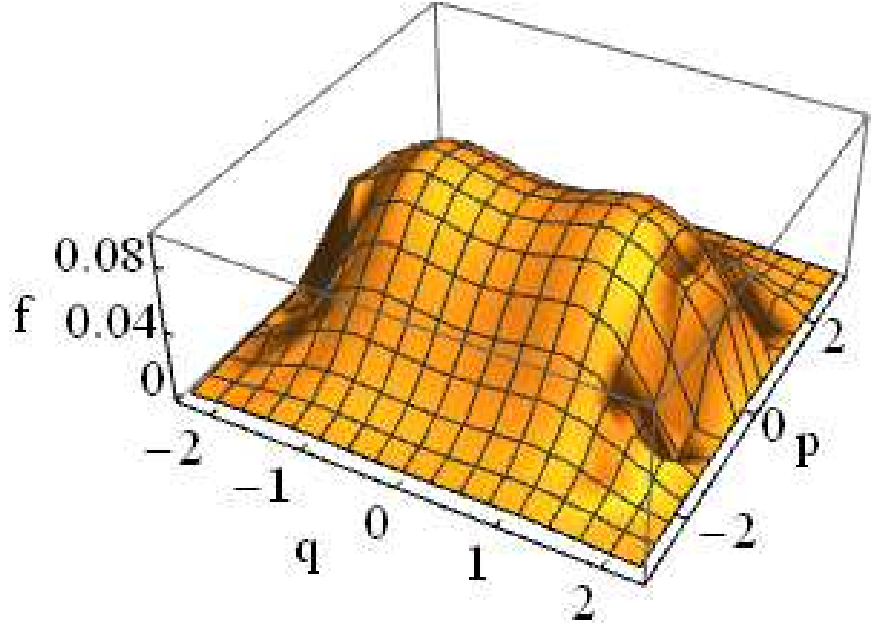}
\caption{Wigner function calculated up to ${\cal O}(\hbar^{10})$ for the one-dimensional Goldstone potential in Eq. (\ref{e22}) and 
$\hbar = 0.6$, with $f_0$ given by Eq. (\ref{fd}) with fugacity $z = 1$.} 
\label{figure1}
\end{center}
\end{figure}

For the sake of illustration, we chose a Fermi-Dirac distribution,
\begin{equation}
\label{fd}
f_0 = \frac{1}{\exp(H)/z + 1} \,,
\end{equation}
where $z = \exp(\mu)$ is the fugacity in terms of the dimensionless chemical potential $\mu$. In terms of the degeneracy parameter $\chi = T_F/T$, where $T$ and $T_F$ are the thermodynamic and Fermi temperatures, one has \cite{Pathria}
\begin{equation}
\label{li}
{\rm Li}_{3/2}(-z) = - \,\frac{4 \chi^{3/2}}{3\sqrt{\pi}} \,. 
\end{equation}
Equation (\ref{li}) contains the polylogarithm function ${\rm Li}_{\nu}(-z)$ defined by 
\begin{equation}
{\rm Li}_{\nu}(-z) = - \frac{1}{\Gamma(\nu)}\int_{0}^{\infty}\frac{s^{\nu-1}\,ds}{\exp(s)/z+1} \,, \quad \nu>0 \,,
\end{equation}
where $\Gamma(\nu)$ is the gamma function. 
In what follows, we set $z = 1$, which corresponds to intermediate degeneracy ($\chi = 1.01$). In addition, in what follows we normalize all Wigner functions to unity ($\int_{\Re} dp \int_{\Re} dq \,f = 1$). In this setting one has the Wigner function shown in Figure \ref{figure1} for $\hbar = 0.6$.

\begin{figure}[h]
\begin{center}
\includegraphics[width=4.5in]{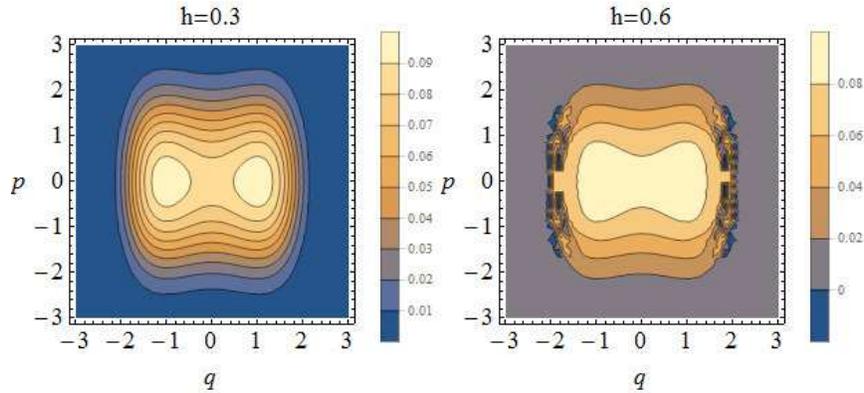}
\caption{Contour plots of the Wigner function with the same choices of Fig. \ref{figure1}, 
for $\hbar = 0.3$ (left) and $\hbar = 0.6$ (right).} 
\label{figure2}
\end{center}
\end{figure}

\begin{figure}[h]
\begin{center}
\includegraphics[width=4.5in]{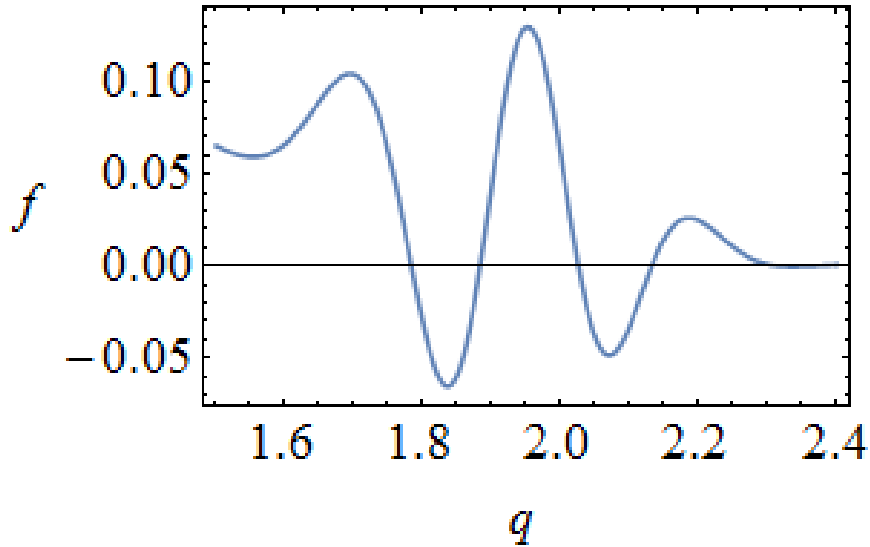}
\caption{Detail of negative values of the Wigner function with the same setting of Figures \ref{figure1} and \ref{figure2}, with $\hbar = 0.6$.}
\label{figure3}
\end{center}
\end{figure}

Figure \ref{figure2} shows the Wigner function contour plots for different values of $\hbar$. It is apparent that for larger quantum effects the fixed points at $q = \pm 1, p = 0$ start to merge due to tunneling, besides showing some negative value regions. 
Negative values of the Wigner function can be also precisely detected, as shown in Figure \ref{figure3}.

\begin{figure}[h]
\begin{center}
\includegraphics[width=4.5in]{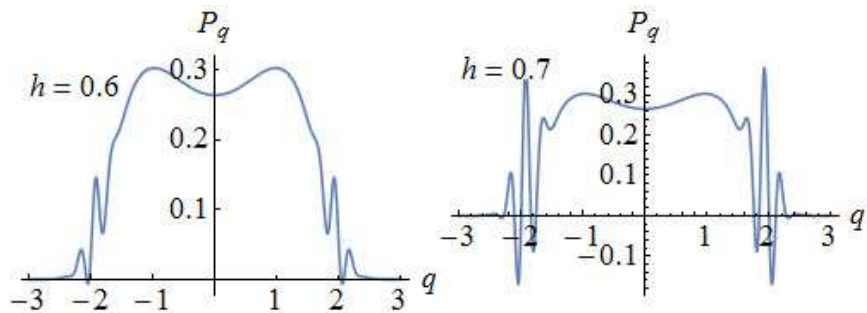}
\caption{Probability distribution $P_q(q,h)$ in position space from Eq. (\ref{x1}), for $\hbar = 0.6$ (left) and $\hbar = 0.7$ (right), using the same settings of the previous figures.} 
\label{figure4}
\end{center}
\end{figure}

\begin{figure}[h]
\begin{center}
\includegraphics[width=4.5in]{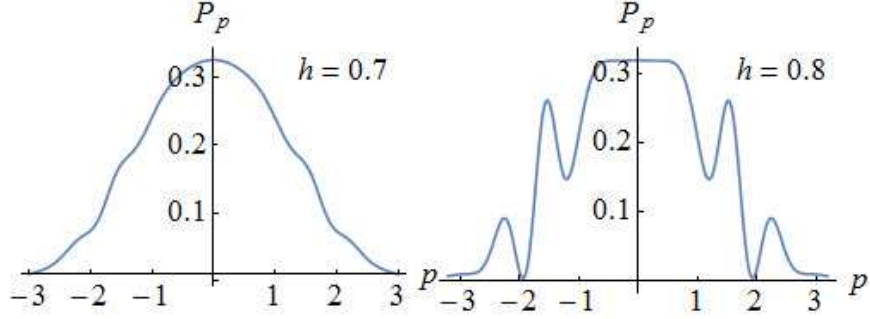}
\caption{Probability distribution $P_p(p,h)$ in position space from Eq. (\ref{x2}), for $\hbar = 0.7$ (left) and $\hbar = 0.8$ (right), using the same settings of the previous figures.} 
\label{figure5}
\end{center}
\end{figure}

The probability distribution $P_q(q,h)$ in position space from Eq. (\ref{x1}) is depicted in Figure \ref{figure4}. As apparent, a larger quantum parameter produces significant regions of negative values of $P_q(q,h)$, which is indicative that the series expansion solution is not sufficiently accurate for such large values of $\hbar$. On the other hand, the appearance of negative values of the probability distribution in momentum space $P_p(p,\hbar)$ from Eq. (\ref{x2}) is not an issue, at least in the present example, as seen in Figure \ref{figure5}.  However, a large $\hbar$ yields a significant distortion and oscillatory pattern of the otherwise Gaussian-like form. 

\begin{figure}[h]
\begin{center}
\includegraphics[width=4.5in]{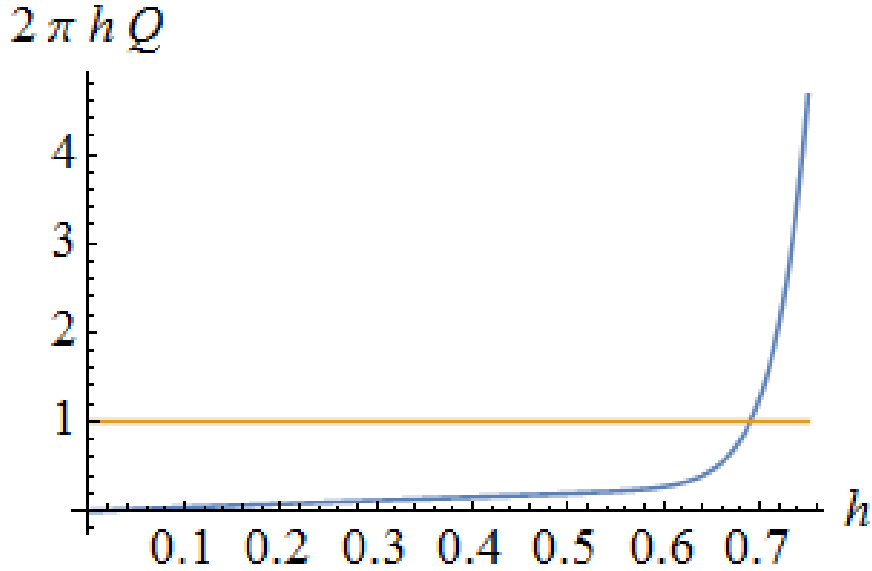}
\caption{Quantity $2\pi\hbar Q(\hbar)$ from Eq. (\ref{x3}) as a function of the quantum diffraction parameter, using the same settings of the previous figures.} 
\label{figure6}
\end{center}
\end{figure}

The quantity $2\pi\hbar Q(\hbar)$ is shown in Figure { \ref{figure6}}, using Eq. (\ref{x3}). For large quantum diffraction parameter one has a growing $Q(h)$ and one verifies that for large $\hbar$ the inequality (\ref{x3}) is not meet anymore, which indicates a violation of the uncertainty principle. This could be expected since in this case a semiclassical expansion would be inappropriate. 

\section{Example: quartic potential without symmetry breaking} 

\begin{figure}[h]
\begin{center}
\includegraphics[width=4.5in]{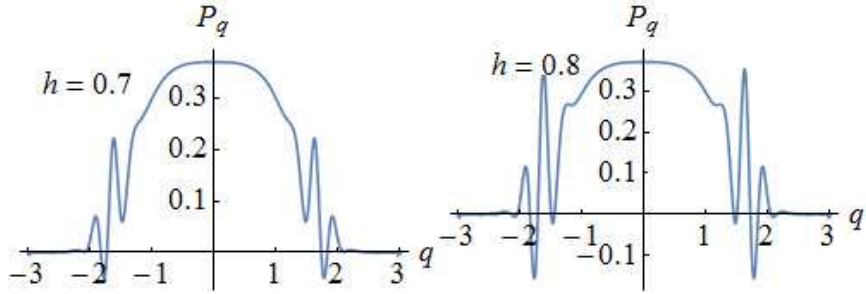}
\caption{Probability distribution $P_q(q,h)$ in position space from Eq. (\ref{x1}), for $\hbar = 0.7$ (left) and $\hbar = 0.8$ (right), for the Fermi-Dirac classical distribution in Eq. (\ref{fd}) with fugacity $z = 1$ and the quartic potential from Eq. (\ref{qua}).} 
\label{figure7}
\end{center}
\end{figure}

We briefly consider the case of a quartic potential without symmetry breaking, 
\begin{equation}
V = \frac{q^4}{4} \,.
\label{qua}
\end{equation}
It obviously belongs to the same class of quartic potentials of the previous example, but with $q = 0$ as the unique stable fixed point. Moreover, there are significant differences for the computer algebra running time and convergence, as separately verified. Omitting the details and performing the quadratures up to ${\cal O}(\hbar^{10})$ with the Fermi-Dirac distribution in Eq. (\ref{fd}) and $z = 1$, we find for instance Figure \ref{figure7} for the marginal probability distribution in configuration space and different quantum diffraction strengths. { The checking of the inequality (\ref{x3}) produces similar results as shown in Figure \ref{figure6}.}

%

% 
%\begin{figure}
%\onefigure{fig8.eps}
%\caption{Quantity $2\pi\hbar Q(\hbar)$ as a function of the quantum diffraction parameter, from Eq. (\ref{x3}) and using the same settings of Figure \ref{figure7}.} 
%\label{figure8}
%\end{figure}
%

\section{Example: confining potential with ripples}

\begin{figure}[h]
\begin{center}
\includegraphics[width=4.5in]{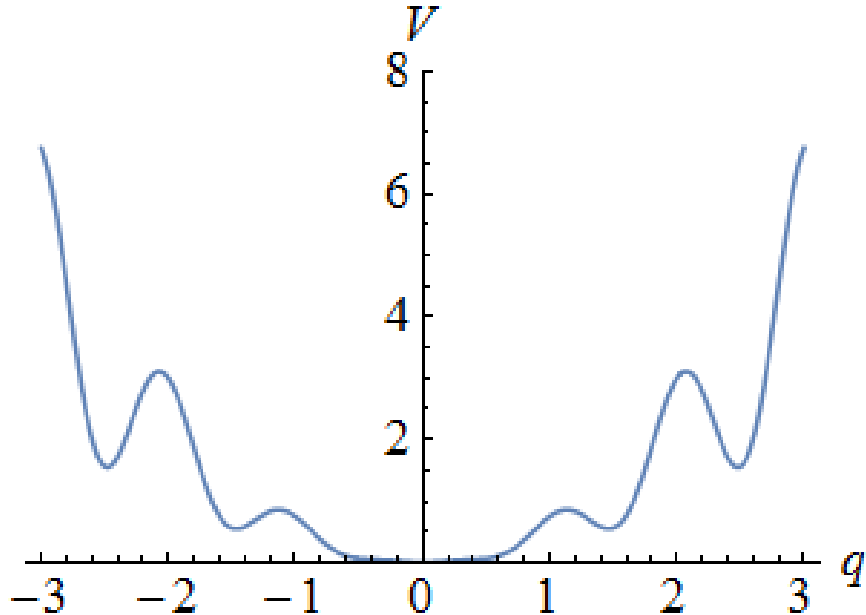}
\caption{Modulated harmonic potential from Eq. (\ref{modu}) with $a = 1$.} 
\label{figure8}
\end{center}
\end{figure}

As a final example, we consider 
\begin{equation}
\label{modu}
V = \frac{q^2}{2}[1 + a \cos(2\pi q)] \,, 0 < a < 1 ,,
\end{equation}
which is a modulated harmonic potential shown in Figure \ref{figure8}. We have performed the series expansion up to ${\cal O}(\hbar^{10})$
with the Fermi-Dirac defined in Eq. (\ref{fd}) with $z = 1$ and the modulation parameter $a = 1/2$. { The results are similar to the previous examples, but it can be verified that the existence of ripples makes the series approximation less efficient already at smaller values of the scaled Planck constant,  restricted to $\hbar < 0.5$ in this case.}

%

% 
%\begin{figure}
%\onefigure{fig10.eps}
%\caption{Probability distribution $P_q(q,h)$ in position space from Eq. (\ref{x1}), for $\hbar = 0.3$ (left) and $\hbar = 0.5$ (right), for the Fermi-Dirac classical distribution in Eq. (\ref{fd}) with fugacity $z = 1$ and the potential from Eq. (\ref{modu}) with $a = 1/2$.} 
%\label{figure10}
%\end{figure}
%

% 
%\begin{figure}
%\onefigure{fig11.eps}
%\caption{Quantity $2\pi\hbar Q(\hbar)$ as a function of the quantum diffraction parameter, from Eq. (\ref{x3}) and using the same settings of Figure \ref{figure11}.} 
%\label{figure11}
%\end{figure}
%

\section{Conclusions}

The main result of this work is Eq. (\ref{e19}), determining the expansion functions $f_l(x,H) \,,\, l = 0,1,2...$ in Eq. (\ref{e15}) by means of a sequence of quadratures once the lower order expansion functions are known. This is always possible in terms of the recursive application of Eq. (\ref{e19}). Therefore we provide a recipe for the almost immediate solution of the stationary one-dimensional Wigner-Moyal equation up to arbitrary order on { quantum-diffraction} effects measured by a dimensionless Planck constant. Obviously the possibility of quick quadratures in configuration space only, is an enormous advancement in comparison with solving a chain of non-trivial partial differential equations at each order \cite{Wigner}. For higher-order quantum-tunneling-effects, the procedure is best carried on using a { 
computer-algebra} package.  An arbitrary classical limit $f_0(H)$ of the Wigner function is allowed, although certain choices can certainly deserve more computer time than others.  Accuracy tests for the series expansion were also provided. The examples of { a one-dimensional} Goldstone potential, of a purely growing quartic potential and of a modulated harmonic confinement have been worked out up to ${\cal O}(\hbar^{10})$. For the sake of definiteness the focus was on the Fermi-Dirac distribution, although the details of { $f_{0}(H)$}  are not decisive for the efficiency of the method.

The r\^ole of the energy integral was for the first time shown in detail, to be decisive for the expedite power series solution of the basic quantum kinetic equation for the Wigner function in a stationary external potential. We provided a formal solution of the quantum Vlasov equation, as a recursive chain of quadratures in position space. In the present context, the Hamiltonian was the appropriate dynamical variable, due to the one-dimensional stationary character. The results are important whenever an accurate Wigner function is necessary, beyond the lowest-order semiclassical ${\cal O}(\hbar^2)$ approximation. The procedure applies for external potentials only.  The case involving a 
self-consistent piece as in ultra-small semiconductor devices and quantum plasmas described by the Wigner-Poisson system \cite{Haas} needs further considerations, since in this situation the potential must obviously be also expanded as a power series on the quantum diffraction parameter.

\acknowledgments
The author acknowledges the support by Con\-se\-lho Na\-cio\-nal de De\-sen\-vol\-vi\-men\-to Cien\-t\'{\i}\-fi\-co e Tec\-no\-l\'o\-gi\-co
(CNPq). %There is no data associated with this article.


\begin{thebibliography}{0}
	
	\bibitem{Wigner} Wigner E. P, Phys. Rev. {\bf 40} (1932) 749.

\bibitem{Weinbub} 
Weinbub J. and Ferry D. K., Appl. Phys. Rev. {\bf 5} (2018) 041104.
	
	\bibitem{Kull}
	Schmidt-Bleker A., Gassen W. and Kull H.-J., Europhys. Lett. {\bf 95} (2011) 55003.
	
	\bibitem{Jung}
	Kim H. M. and Jung Y. D., Europhys. Lett. {\bf 78} (2007) 35001.
	
	\bibitem{Boni}
	Bonifacio R., Cola M. M., Piovella N. and Robb G. R. M., Europhys. Lett. {\bf 69} (2005) 55.
	
	\bibitem{Else} 
	Else D., Kompaneets R. and  Vladimirov S. V., Europhys. Lett. {\bf 94} (2011) 35001.
	
	\bibitem{Jungel2009} 
  J\"ungel A., Transport Equations for Semiconductors.  
 % \Editor{A. Editor}
 % \Vol{9}
  Springer, Berlin-Heidelberg, 2009.
  %\Page{666}.

\bibitem{Bernstein} 
Bernstein I. B., Greene J. M. and Kruskal M. D., Phys Rev. {\bf 108} (1957) 546.

\bibitem{Luque} 
Luque A., Schamel H. and Fedele R., Phys. Lett. A {\bf 324} (2004) 185.

\bibitem{Demeio} 
Demeio L., Transp. Theory Stat. Phys. {\bf 36} (2007) 137.

\bibitem{Smerzi} %A. Smerzi, Phys. Rev. A {\bf 52}, 4365 (1995).
Smerzi A., Phys. Rev. A {\bf 52} (1995) 4365.

\bibitem{Lange} %H. Lange, B. Toomire, and P. F. Zweifel, Transp. Theor. Stat. Phys. {\bf 25}, 713 (1996).
Lange H., Toomire B. and Zweifel P. F., Transp. Theor. Stat. Phys. {\bf 25} (1996) 713.

\bibitem{Bose} %A. Bose, and M. S. Janaki, Eur. J. Phys. B {\bf 87}, 259 (2014). 
Bose A. and Janaki M. S., Eur. J. Phys. B {\bf 87} (2014) 259.

\bibitem{HS} %F. Haas, and P. K. Shukla, Phys. Plasmas {\bf 15}, 112302 (2008).
Haas F. and Shukla P. K., Phys. Plasmas {\bf 15} (2008) 112302.

\bibitem{Schuch} %D. Schuch, and M. Moshinsky, Phys. Rev. A {\bf 73}, 062111 (2006).
Schuch D. and Moshinsky M., Phys. Rev. A {\bf 73} (2006) 062111.

\bibitem{NJP} %F. Haas, J. Zamanian, M. Marklund, and G. Brodin, New J. Phys. {\bf 12}, 073027 (2010).
Haas F., Zamanian J., Marklund M. and Brodin G., New J. Phys. {\bf 12} (2010) 073027.

\bibitem{Nedjalkov} %M. Nedjalkov, J. Weinbub, M. Ballicchia, S. Selberherr, I. Dimov, and D. K. Ferry, Phys. Rev. B {\bf 99}, 014423 (2019).
Nedjalkov M., Weinbub J., Ballicchia M., Selberherr S., Dimov I. and Ferry D. K., Phys. Rev. B {\bf 99} (2019) 014423.

\bibitem{Taylor} %M. Taylor, {\it Pseudo-differential Operators} (Princeton University Press, Princeton,1981).
  Taylor M., Pseudo-differential Operators. 
  %\Editor{A. Editor}
  %\Vol{9}
  Princeton University Press, Princeton, 1981.
  %\Page{666}.

\bibitem{Markowich} %P. A. Markowich, C. A. Ringhofer, and C. Schmeiser, {\it Semiconductor Equations} (Springer, Wien, 1990).
 Markowich P. A., Ringhofer C. A. and Schmeiser C., Semiconductor Equations. 
  %\Editor{A. Editor}
  %\Vol{9}
  Springer, Wien, 1990.
  %\Page{666}.


\bibitem{Hillery} %M. Hillery, R. F. O'Connell, M. O. Scully, and E. P. Wigner, Phys. Rep. {\bf 106}, 121 (1985).
Hillery M., O'Connell R. F., Scully M. O. and Wigner E. P., Phys. Rep. {\bf 106} (1985) 121.

\bibitem{Manfredi} %G. Manfredi, and M. R. Feix, Phys. Rev. E {\bf 53}, 6460 (1996).
Manfredi G. and Feix M. R., Phys. Rev. E {\bf 53} (1996) 6460.

\bibitem{Pathria} 
  Pathria R. K. and Beale P. D., Statistical Mechanics 3rd ed. 
  %\Editor{A. Editor}
  %\Vol{9}
  Academic Press, Cambridge, 2011, p. 182.

\bibitem{Haas} %F. Haas, {\it Quantum Plasmas: an Hydrodynamic Approach} (Springer, New York, 2011). 
  Haas F., Quantum Plasmas: an Hydrodynamic Approach
  %\Editor{A. Editor}
  %\Vol{9}
  Springer, New York, 2011.
  %\Page{666}.

\end{thebibliography}
\end{document}